\documentclass[%
 aip,
 amsmath,amssymb,
 reprint,%
]{revtex4-1}

\usepackage{graphicx}
\usepackage{dcolumn}
\usepackage{bm}

\usepackage[utf8]{inputenc}
\usepackage[T1]{fontenc}
\usepackage{mathptmx}
\usepackage{etoolbox}
\usepackage{xcolor}

\makeatletter
\def\@email#1#2{%
 \endgroup
 \patchcmd{\titleblock@produce}
  {\frontmatter@RRAPformat}
  {\frontmatter@RRAPformat{\produce@RRAP{*#1\href{mailto:#2}{#2}}}\frontmatter@RRAPformat}
  {}{}
}%
\makeatother
\begin{document}

\preprint{AIP/123-QED}
\newcommand{\tex}{\textcolor{black}}

\title[Deterministic Creation of Strained Color Centers in Nanostructures via High-Stress Thin Films]{Deterministic Creation of Strained Color Centers in Nanostructures via High-Stress Thin Films}
\author{D. R.  Assumpcao}
\author{C. Jin}%
\affiliation{ 
John A. Paulson School of Engineering and Applied Sciences, Harvard University, Cambridge, MA, 02138, USA
}%

\author{M. Sutula}

\affiliation{ 
Department of Physics , Harvard University, Cambridge, MA, 02138, USA
}%

\author{S. W. Ding}
\affiliation{ 
John A. Paulson School of Engineering and Applied Sciences, Harvard University, Cambridge, MA, 02138, USA
}%

\author{P. Pham}
\affiliation{ 
John A. Paulson School of Engineering and Applied Sciences, Harvard University, Cambridge, MA, 02138, USA
}%

\author{C. M. Knaut}
\affiliation{ 
Department of Physics , Harvard University, Cambridge, MA, 02138, USA
}%

\author{M. K. Bhaskar}
\affiliation{ 
Department of Physics , Harvard University, Cambridge, MA, 02138, USA
}%
\affiliation{ 
AWS Center for Quantum Networking, Boston, MA 02135, USA
}%

\author{A. Panday}
\affiliation{ 
John A. Paulson School of Engineering and Applied Sciences, Harvard University, Cambridge, MA, 02138, USA
}%

\author{A. M. Day}
\affiliation{ 
John A. Paulson School of Engineering and Applied Sciences, Harvard University, Cambridge, MA, 02138, USA
}%

\author{D. Renaud}
\affiliation{ 
John A. Paulson School of Engineering and Applied Sciences, Harvard University, Cambridge, MA, 02138, USA
}%

\author{M. D. Lukin}
\affiliation{ 
Department of Physics , Harvard University, Cambridge, MA, 02138, USA
}%

\author{E. Hu}
\affiliation{ 
John A. Paulson School of Engineering and Applied Sciences, Harvard University, Cambridge, MA, 02138, USA
}%

\author{B. Machielse}
\affiliation{ 
Department of Physics , Harvard University, Cambridge, MA, 02138, USA
}%
\affiliation{ 
AWS Center for Quantum Networking, Boston, MA 02135, USA
}%

\author{M. Loncar}
\affiliation{ 
John A. Paulson School of Engineering and Applied Sciences, Harvard University, Cambridge, MA, 02138, USA
}%
 \email{loncar@seas.harvard.edu}

\date{\today}

\begin{abstract}
Color centers have emerged as a leading qubit candidate for realizing hybrid spin-photon quantum information technology. One major limitation of the platform, however, is that the characteristics of individual color-centers are often strain dependent. As an illustrative case, the silicon-vacancy center in diamond typically requires millikelvin temperatures in order to achieve long coherence properties, but strained silicon vacancy centers have been shown to operate at temperatures beyond 1K without phonon-mediated decoherence. In this work we combine high-stress silicon nitride thin films with diamond nanostructures to reproducibly create statically strained silicon-vacancy color centers (mean ground state splitting of 608 GHz) with strain magnitudes of $\sim 4 \times 10^{-4}$. Based on modeling, this strain should be sufficient to allow for operation of a majority silicon-vacancy centers within the measured sample at elevated temperatures (1.5K) without any degradation of their spin properties. This method offers a scalable approach to fabricate high-temperature operation quantum memories. Beyond silicon-vacancy centers, this method is sufficiently general that it can be easily extended to other platforms as well.
\end{abstract}

\maketitle

\begin{figure}
\includegraphics{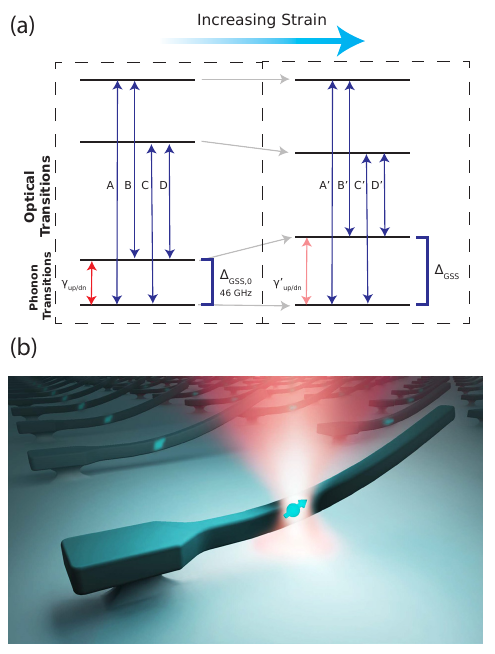}
\caption{\label{fig:overview} Combining strained thin films with diamond nanostructures to deterministically strain silicon-vacancy centers (SiV). (a) Level diagram of the SiV electron spin. Phonon mediated transitions ($\gamma_{up/dn}$) between the bottom two energy levels ($\Delta_{GSS}$) is the primary dephasing mechanism for the SiV at elevated temperatures (> ~100 mK). Upon experiencing strain perpendicular to the SiV's internal Z axis, however, a significant shift in the SiV’s energy levels leads to a shift in optical transition frequencies and therefore a significant increase in $\Delta_{GSS}$. This suppresses $\gamma_{up/dn}$ which enables higher temperature operation of the color center without spin coherence degradation. (b) Schematic of the diamond-SiN structure whereby SiN is deposited on diamond cantilevers to introduce significant strain into the cantilever and thereby deterministically strain the SiVs. }
\end{figure}

Solid-state color-center defects are an exciting platform for quantum information technology which combine scalable fabrication, state-of-the-art spin properties, and excellent photonic interfaces \cite{Bradac2019, Ruf2021, Castelletto2020}. This combination enables the realization of interfaces between long-lived memories and individual optical photons for a variety of applications including quantum communication, computation, and sensing \cite{Tiurev2022, Castelletto2020, Ruf2021, Kimble2008}. Recently the silicon-vacancy (SiV) center in diamond has emerged as a leading color center defect primarily due to its inversion symmetry which enables integration of SiVs into nanostructures without a significant degradation of the defect's spin or optical properties \cite{Evans2016}. Pioneering work has utilized SiVs implanted within a diamond nanophotonic resonator to perform a variety of demonstrations including deterministic spin-photon interactions, high-efficiency single photon generation, and memory-enhanced quantum communication \cite{Nguyen2019, Knall2022, Bhaskar2020}. 

One major challenge associated with color-center qubits is their sensitivity to their local environment. For example, the local strain environment of the defect can have a significant impact on the properties of the quantum emitter \cite{Castelletto2020, Meesala2018, Maity2018, Falk2014, Knauer2020, Batalov2009, Olivero2013, Udvarhelyi2018, Li2020, Lindner2018}. In the case of the SiV, one major limitation of the emitter is its extreme sensitivity to phonon-induced decoherence via driving of the transition between the upper and lower orbital branches of its ground state manifold (Fig.~\ref{fig:overview}(a)) \cite{Meesala2018}. The splitting between these transitions, known as the ground state splitting ($\Delta_{GSS}$), is ~46 GHz for an unstrained SiV which necessitates operation at millikelivin temperatures to freeze out higher-energy, resonant phonons and obtain long coherence properties \cite{Sukachev2017}. 

Alternatively, recent work has shown that strained SiVs can overcome this limitation through a strain-induced enlargening of $\Delta_{GSS}$, thus requiring higher-energy phonons to decohere the spin and therefore enabling higher-temperature operation \cite{Meesala2018, Sohn2018}. Previous work has already shown that the coherence of an SiV at 4K can be improved through dynamically straining the SiV via active MEMS tuning \cite{Meesala2018, Sohn2018}. More recently, a strained SiV was shown to be operable up to 1.5K without any phonon-induced degradation of its spin properties, achieving coherence times beyond 100 $\mu s$ \cite{Stas2022}. Additionally, strain has also been shown to improve other properties of the SiV including its spectral stability and microwave susceptibility of its spin for coherent control \cite{Nguyen2019}.  Despite the clear benefits of utilizing strained SiVs, a simple, general method to deterministically strain the static environment of the defect within a nanostructure is lacking. Previous work utilized either complex nano-electro-mechanical  tuning, the random distribution of strain in the nanofabricated structures which is inherently probabilistic and has a low yield, or thermal-streses induced within a diamond thin film which is platform-specific and has yet to be effectively translated to strain within nanostructures  \cite{Meesala2018, Machielse2019, Stas2022, guo2023microwavebased}.

\begin{figure*}
\includegraphics{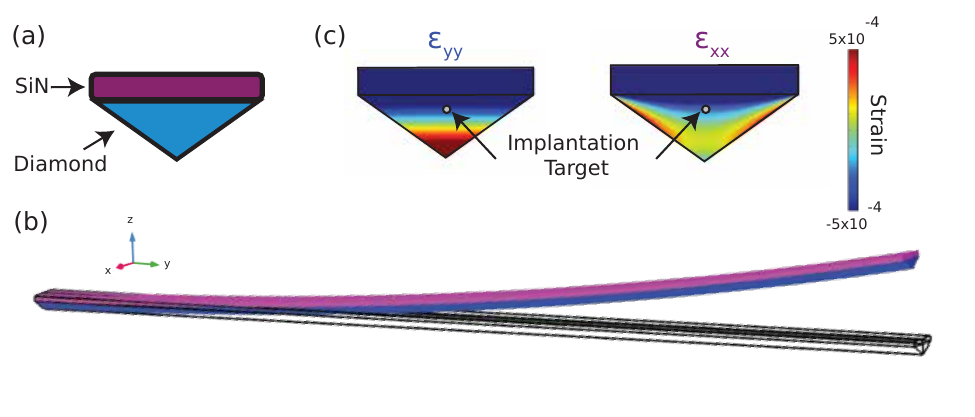}
\caption{\label{fig:simulation} Design of SiN-diamond heterostructure. (a). Cross-sectional schematic of the device, consisting of an intrinsically stressed SiN thin-film on a free standing diamond cantilever. A triangular cross-section diamond device fabricated via reactive ion-beam angled etching is used \cite{Chia2022}.  (b) FEM simulations of SiN-diamond cantilevers, showing significant strain is imparted into the nanostructure through utilization of a high-intrinsic stress SiN thin film. The magnitude of stress of the thin film considered in the simulations is based on the estimated thin film stress at 4K based on the experimentally measured SiV distributions (see Fig \ref{fig:coherence}). (c) Cross-sectional strain profile of the cantilevers taken at the center of the cantilever showing significant axial strain (> $1 \times 10^{-4}$) is present at the target implantation point of the SiVs parallel and in-plane perpendicular to the beam. We note that although the profile is taken from the center of the cantilever, the strain within the beam is quite uniform along the beam. }
\end{figure*}

In this work, we propose a method for reproducibly creating strained SiVs through integrating intrinsically strained thin film stressors with diamond nanostructures. We integrate high stress silicon-nitride (SiN) thin films with diamond cantilevers containing SiVs (Fig.~\ref{fig:overview}(b)). We measure the optical properties of the SiVs within these nanostructures before and after SiN deposition and observe a significant increase in the average $\Delta_{GSS}$ up to 608 GHz, \tex{over 5 times greater than the average $\Delta_{GSS}$ in the sample prior to SiN deposition and thus indicative of} the introduction of significant strain. Through modeling based on previous published results, we estimate this splitting is sufficient to enable 1.5K operation for a majority of SiVs in the sample. Our approach is therefore a suitable method for reproducibly creating highly-strained SiVs in nanostructures.

For our design we consider a freestanding diamond cantilever aligned to the [110] direction of the diamond with a 60 nm thin film of tensile-stressed SiN on the top surface and SiVs integrated via ion implantation (Fig.~\ref{fig:simulation}(a)) \cite{Nguyen2019}. SiN is selected as the stressor due to the high intrinsic stress achievable with the material, ease of deposition, and its previous use as a stressor for other integrated photonics applications \cite{Jacobsen2006, Ke2017, Ghrib2013, Sukhdeo2015}. Tensile stress is chosen so that the additional thermal stress between the SiN film and diamond cantilever upon cooling down, induced by a mismatch in thermal expansion coeffficients, adds constructively to this intrinsic stress. 

In order to estimate the distribution of expected strains, Finite-Element-Method (FEM, COMSOL) simulations are performed. We observe a significant strain in the cantilever (Fig.~\ref{fig:simulation}(b)). The axial strain components located parallel  ($\epsilon_{yy}$) and perpendicular in-plane ($\epsilon_{xx}$) relative to the cantilever (Fig.~\ref{fig:simulation}(c)) are dominant and strongly depth dependent. To account for this, we target an SiV implantation depth of 35 nm as a trade-off between maximizing the strain experienced by the defect and ensuring they are far enough from the surface to not experience potentially detrimental surface charge noise. The biaxial nature of the strain ensures that the SiV will experience a strain perpendicular to its internal Z-axis of sufficient magnitude to significantly increase $\Delta_{GSS}$ regardless of its orientation within the crystal \cite{Meesala2018}.

\begin{figure*}
\includegraphics{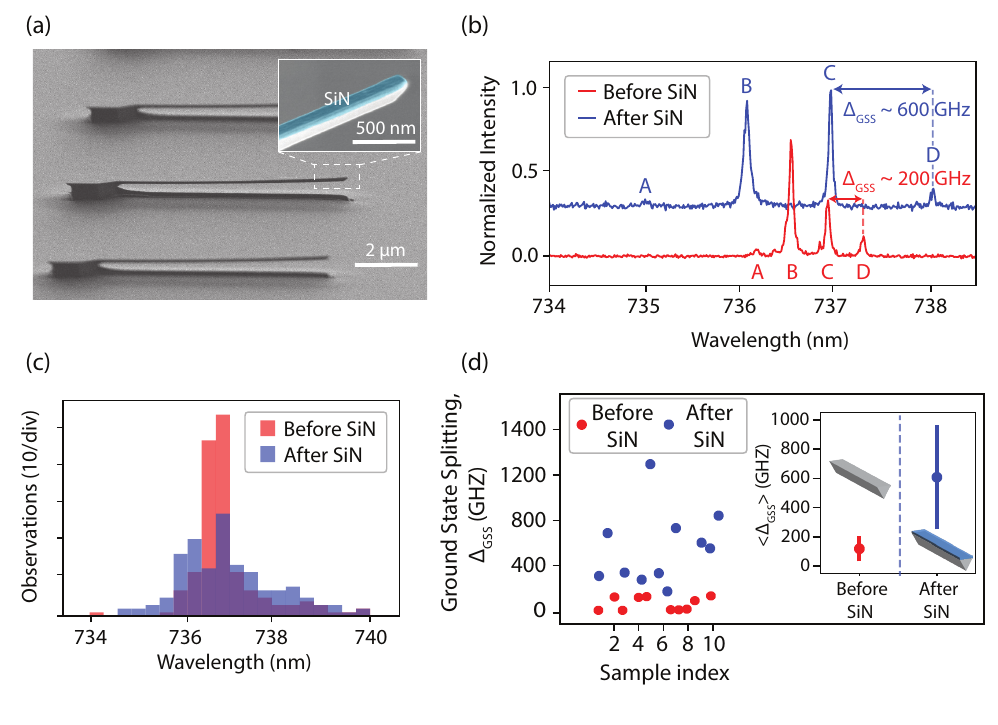}
\caption{\label{fig:experiment} Experimental verification of strained SiV formation (a) Scanning electron micrograph (SEM) of a diamond cantilever with high-stress SiN deposited, showing a significant strain imparted on the cantilever. (b). Photoluminescence spectra of the same SiV at 4K before (red) and after (blue) deposition of the SiN, showing a shift in all four optical lines and an increase in the $\Delta_{GSS}$ of the SiV due to the imparted strain. (c) Histogram of all observed SiV transition locations before and after SiN deposition, showing a significant increase in the distribution of transition locations upon deposition of nitride which is indicative of strain. (d) Measured $\Delta_{GSS}$ for a subset of SiVs before and after the deposition of SiN, showing a large increase in $\Delta_{GSS}$ upon the deposition of SiN as desired. (inset) Average $\Delta_{GSS}$ and the \tex{standard deviation} for the distributions of measured SiVs before and after SiN deposition.  }
\end{figure*}

Having numerically shown the ability for stressed SiN films to impart significant strain in diamond cantilevers, corresponding samples are fabricated. Free-standing diamond cantilevers are fabricated using angled ion-beam etching. SiVs are precisely formed within the cantilever through combining  masked ion-implantation of Si atoms (50 keV) with high temperature annealing at 1250 \textdegree C in ultra-high vacuum \cite{Nguyen2019}. The density of the SiVs is kept low such that individual SiV centers are spectrally resolvable. After device fabrication, a 60 nm thin-film of SiN is deposited using plasma-enhanced chemical-vapor deposition (PECVD). The low deposition temperature required by PECVD ensures process compatibility both with the diamond nanostructures used in this work and potentially a variety of other quantum photonic platforms \cite{Balseanu2006}. A significant bending is observed in the cantilever upon SiN deposition, indicative of thin-film induced strain in the nanostructure (Fig.~\ref{fig:experiment}(a)). 

In order to observe the effects of the SiN stressor on the SiVs, off-resonance excitation based photoluminescence (PL) spectra are taken of the SiVs at 4K before and after deposition of the SiN. A significant shift of the spectral lines is observed after the SiN is deposited which we attribute to the strain induced by the SiN (Fig.~\ref{fig:experiment}(b)). Specifically we observe an increased separation between the individual transitions which is indicative of an increasing of the $\Delta_{GSS}$ as expected and desired. To better quantify this, PL spectra from approximately 100 implantation spots are taken and individual peaks in each spectra are located corresponding with any of the four SiV optical transitions. Upon deposition of the SiN, we see a significant broadening in the distribution of the peaks, again indicating a shift in the optical lines that we attribute to strain induced by the SiN (Fig.~\ref{fig:experiment}(c)). We note that this distribution does not differentiate between the different SiV optical lines and thus provides information on the distribution of all the optical transitions. As further verification that strain induced by the SiN stressor is indeed responsible for this broadening, spectra are taken from SiVs implanted at the base of the cantilever where less stress is expected due to the mechanical tether to the substrate, and no significant broadening is observed (see Supporting Information). \tex{We note that although we observe a change in the relative intensity of the different optical transitions upon SiN deposition (Fig.~\ref{fig:experiment}(b)), we do not observe a significant change in total PL counts after SiN deposition, indicating the defect's optical properties are not degraded by the presence of the thin film.}

In order to better quantify the magnitude of strain observed, we identify PL spectra where four or fewer clear lines are observed (11 spectra out of the 100 measured), which we interpret as being from an individual SiV. From these spectra, we assume the two lowest-energy transitions correspond to the C and D transitions of the SiV respectively, and thus the difference in frequencies of the transitions corresponds to $\Delta_{GSS}$ of the SiV. From this we are able to extract $\Delta_{GSS}$ for a handful of SiVs before and after SiN deposition (Fig.~\ref{fig:experiment}(d)). We extract the mean and standard error of the mean from this set of measured $\Delta_{GSS}$ to be 119 $\pm$ 22 GHz prior to SiN deposition and 608 $\pm$ 89 GHz after deposition, demonstrating a significant overall increase in $\Delta_{GSS}$.  A > 5x increase of the mean of $\Delta_{GSS}$ indicates strains on the order of $4 \times 10^{-4}$ within the nanostructure. We thus have verified the ability to realize strained SiVs reproducibly across the sample through integration of strained thin films with nanostructures.

\begin{figure}
\includegraphics{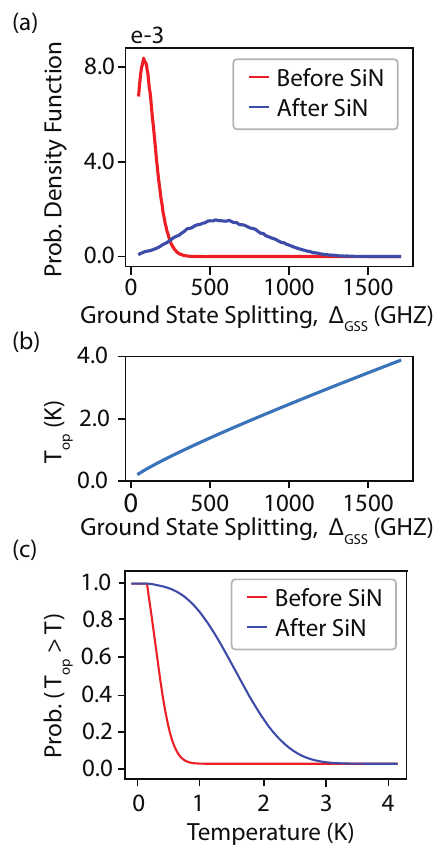}
\caption{\label{fig:coherence} Modeling the coherence enhancement of strained SiVs. (a) Modeled probability density function for $\Delta_{GSS}$ of the SiVs in the cantilever before and after deposition of SiN, showing a significant increase in probability of larger $\Delta_{GSS}$ splittings in the SiN case. (b) Estimated operational temperature ($T_{op}$) of an SiV with a given $\Delta_{GSS}$ without degradation of coherence properties, demonstrating that an increased $\Delta_{GSS}$ significantly enhances an SiV’s operational temperature. (c) Probability of finding an SiV with an operational temperature ($T_{op}$) above a given temperature $T$, before and after SiN deposition. After SiN deposition, we see a significant enhancement in the probability of finding an SiV operational at 1.5K and beyond, thus highlighting that sufficient strain is generated in the nanostructure  by the thin film to enhance coherence properties at elevated temperatures. }
\end{figure}

To better understand the potential benefits of thin film induced static strain for SiVs, we combine our measured experimental results with modeling to extrapolate the high-temperature performance of the SiVs in the nanostructures. As a first step, we extrapolate distributions for the $\Delta_{GSS}$ of the SiVs in the cantilever before and after deposition of the SiN stressor based on the experimental results. For the case of the SiVs prior to SiN deposition, we assume the strain experienced by the SiV will be dominated by random \tex{strain originating from implantation of the defect atoms and the subsequent fabrication of the nanostructure. Because of this the strain will not have a preferred direction, and will instead be randomly oriented. \cite{Evans2016, Meesala2018}}. Thus we model each component of the strain tensor experienced by the SiV as an independent normally distributed random variable with a mean of zero and a standard deviation $\sigma_{unstrained}$. From this random tensor distribution, the expected distribution of $\Delta_{GSS}$ is computed (Fig.~\ref{fig:coherence}(a)). To fit this to our experimental results, we vary $\sigma_{unstrained}$ so that the mean of the simulated $\Delta_{GSS}$ distribution matches the mean of the measured distribution, yielding $\sigma_{unstrained} = 1.9 \times 10^{-5}$. As validation of the model, the measured standard deviation of the $\Delta_{GSS}$ distribution of 68 GHz  closely matches the simulated standard deviation of 52 GHz. \tex{Additionally we note that the extracted C-line distribution from this model (standard deviation of 45 GHz) matches closely the measured distribution of SiV C-lines under similar fabrication procedures (standard deviation of 31 GHz), indicating comparative magnitudes of intrinsic strain have been previously observed \cite{Meesala2018}}.

For the case of post-SiN deposition, we assume the strain profile of the cantilever is dominated by the SiN-induced strain and thus is accurately captured by our FEM numerical model. We assume a random distribution of SiV positions given by the combination of the straggle of the Si atom upon ion implantation (estimated using Stopping-Range-in-Matter simulations) and the dimensions of the mask aperture (60 nm by 60 nm). A random sampling of strains experienced by the SiV is then polled based on this random position distribution and the FEM simulation. From this strain distribution, the resulting SiV $\Delta_{GSS}$ distribution is then computed for the different SiV orientations (Fig.~\ref{fig:coherence}(b)). In order to fit this to the experimental results, the initial intrinsic stress of the SiN thin film in the FEM simulation is scaled so that the mean of the simulated $\Delta_{GSS}$ distribution matches the measured mean $\Delta_{GSS}$ in the nanostructures. This yields an equivalent thin film stress of 700 MPa of the SiN thin film, which we note is likely a combination of the intrinsic stress of the SiN and the thermally induced stress due to its differing thermal expansion coefficient from diamond. We again observe a close correspondence between the measured standard deviation of the $\Delta_{GSS}$ distribution of 295 GHz and the simulated standard deviation of 249 GHz, providing validation of the approach. As expected, the modeled distribution of $\Delta_{GSS}$ with SiN has a significantly higher proportion of SiVs with a large $\Delta_{GSS}$ than without SiN.

 Finally, based on the modeled enhancement in $\Delta_{GSS}$, we estimate the improvement in the high-temperature coherence properties of the strained SiVs. We base our analysis on the results of Ref \onlinecite{Stas2022} in which a strained SiV with a ground state splitting of $\Delta_{GSS, 0} = 554$ GHz showed no phonon-induced degradation of the spin coherence properties (with $T_2 > 300 \mu s$) up to a temperature of $T_0 = 1.5$K. The rate of excitation of the SiV electron from the lower branch of the ground state manifold to the upper branch, which is the primary phonon-induced dephasing pathway, is given by $\gamma_{up}(\Delta_{GSS}, T) \propto  (\Delta_{GSS})^3 n_{th}(\Delta_{GSS}, T)$ where $T$ is the temperature of the SiV and $n_{th}$ is the proportion of thermal occupation of the excited branch which is given by the Boltzmann distribution \cite{Meesala2018}. Given the results from Ref \onlinecite{Stas2022}, $\gamma_{up} (\Delta_{GSS, 0}, T_0)$ is sufficiently suppressed by the SiV's $\Delta_{GSS, 0}$ such that it negligibly impacts the coherence. From this we can extrapolate the necessary $\Delta_{GSS}$ for an SiV to be operated at an arbitrary temperature $T_{op}$ with an equivalent suppression of phonon-induced decoherence, $\gamma_{up}(\Delta_{GSS}, T_{op}) = \gamma_{up}(\Delta_{GSS, 0}, T_0)$ and thus an equivalent suppression of phonon-induced decoherence (Fig.~\ref{fig:coherence}(b)). By combining this with our modeled $\Delta_{GSS}$ distributions, we can estimate the probability of finding an SiV within the device which is operable up to a given temperature without phonon-induced decoherence (Fig.~\ref{fig:coherence}(c)).

 We observe that due to the presence of the SiN on the cantilever, a majority of SiVs can be operated beyond 1.5 K. Beyond that, a significant portion (> $20\%$) can be operated beyond 2K. In comparison, the probability of finding an SiV that is sufficiently strained to be operated at these temperatures prior to the deposition of SiN in the measured samples is negligible, highlighting the key role the stressor plays. This indicates that through utilization of this strained thin film, SiVs with sufficient strain to be operated at elevated temperatures beyond 1K are reproducibly created, which substantially relaxes the cryogenic requirements of the platform thereby significantly enhancing its scalability. 

In conclusion, we have demonstrated a method to reproducibly fabricate strained color centers through combining nanostructures with strained SiN thin films. Through utilizing this technique with SiVs in diamond cantilevers, we observe a significant deterministic straining of SiVs with strains on the order of $4 \times 10^{-4}$ and an average $\Delta_{GSS}$ of the emitter of 608 GHz. Based on the observed strains, we expect that the majority of SiVs within this sample can operated at 1.5K and over 20$\%$ at 2K without any significant phonon-induced degradation of coherence properties. These results thus pave the way towards ensuring the scalablility of the state-of-the-art performance achieved with SiV based spin-photon interfaces to realize true large-scale color-center based quantum photonic systems. \tex{Although beyond the scope and cryogenic capabilities utilized in this current work}, future work will probe the coherence properties of these strained SiVs at 1.5K and beyond to demonstrate the benefit of strain. \tex{Furthermore, future work will}  further maximize the strain in the nanostructure through additional process optimizations, investigate the integration of thin-films with dynamically tunable stress \cite{Grim2019},  and explore integration of stressed thin films with other diamond nanostructures, namely nanophotonic cavities, to realize high-yield high-operation temperature spin-photon interfaces. Beyond SiVs, the relative simplicity of the technique and ease of compatibility of deposition of SiN thin films ensures this same technique can be applied to other color centers in both diamond and other material platforms to create a variety of different reproducibly strained defects. 

\tex{
\section*{Supplementary Information}
See the supplementary information for a comparison of SiV transition distributions on the base and on the cantilever proper.
}
\begin{acknowledgments}
This work was supported by AWS Center for Quantum Networking’s research alliance with the Harvard Quantum Initiative, NSF OMA-2137723, EEC-1941583, ONR N00014-20-1-2425, and ARO MURI W911NF1810432. D.R. and D.A. acknowledge support from the NSF GRFP (No. DGE1745303). M.S. acknowledges support from the NASA Space Technology Graduate Research Fellowship Program. D.R. acknowledges support from the Ford Foundation fellowship. P.P. acknowledges support from the Caltech Summer Undergraduate Research Fellowships program. Device fabrication was performed at the Center for Nanoscale Systems (CNS), a member of the National Nanotechnology Coordinated Infrastructure Network (NNCI), which is supported by the National Science Foundation under NSF Grant No. 1541959.

\end{acknowledgments}

\section*{Author Declarations}
\subsection*{Conflict of Interest}
The authors have no conflicts to disclose.
\subsection*{Author Contributions}
\textbf{Daniel Assumpcao}: Data Curation (lead), Formal Analysis (lead), Methodology (equal), Investigation (equal), Writing/Original Draft Preparation (lead). \textbf{Chang Jin}: Investigation (equal). \textbf{Madison Sutula}: Investigation (equal), Formal Analysis (supporting).  \textbf{Sophie Ding} Investigation (supporting). \textbf{Phuong Pham}: Investigation (supporting). \textbf{Can Knaut}: Investigation (supporting). \textbf{Mihir Bhaskar}: Supervision (supporting). \textbf{Abishrant Panday} Investigation (supporting). \textbf{Aaron Day}: Investigation (supporting). \textbf{Dylan Renaud}: Investigation (supporting), Visualization (equal). \textbf{Mikhail Lukin}: Funding Acquisition (supporting). \textbf{Evelyn Hu}: Funding Acquisition (supporting), Supervision (supporting). \textbf{Bartholomeus Machielse}: Supervision (lead), Methodology (equal). \textbf{Marko Loncar}: Supervision (supporting), Project Administration (lead), Funding Acquistion (lead). 

\section*{Data Availability Statement}

The data that support the findings of this study are available from the corresponding author upon reasonable request.


\bibliography{export}

\begin{thebibliography}{30}%
\makeatletter
\providecommand \@ifxundefined [1]{%
 \@ifx{#1\undefined}
}%
\providecommand \@ifnum [1]{%
 \ifnum #1\expandafter \@firstoftwo
 \else \expandafter \@secondoftwo
 \fi
}%
\providecommand \@ifx [1]{%
 \ifx #1\expandafter \@firstoftwo
 \else \expandafter \@secondoftwo
 \fi
}%
\providecommand \natexlab [1]{#1}%
\providecommand \enquote  [1]{``#1''}%
\providecommand \bibnamefont  [1]{#1}%
\providecommand \bibfnamefont [1]{#1}%
\providecommand \citenamefont [1]{#1}%
\providecommand \href@noop [0]{\@secondoftwo}%
\providecommand \href [0]{\begingroup \@sanitize@url \@href}%
\providecommand \@href[1]{\@@startlink{#1}\@@href}%
\providecommand \@@href[1]{\endgroup#1\@@endlink}%
\providecommand \@sanitize@url [0]{\catcode `\\12\catcode `\$12\catcode `\&12\catcode `\#12\catcode `\^12\catcode `\_12\catcode `\%12\relax}%
\providecommand \@@startlink[1]{}%
\providecommand \@@endlink[0]{}%
\providecommand \url  [0]{\begingroup\@sanitize@url \@url }%
\providecommand \@url [1]{\endgroup\@href {#1}{\urlprefix }}%
\providecommand \urlprefix  [0]{URL }%
\providecommand \Eprint [0]{\href }%
\providecommand \doibase [0]{http://dx.doi.org/}%
\providecommand \selectlanguage [0]{\@gobble}%
\providecommand \bibinfo  [0]{\@secondoftwo}%
\providecommand \bibfield  [0]{\@secondoftwo}%
\providecommand \translation [1]{[#1]}%
\providecommand \BibitemOpen [0]{}%
\providecommand \bibitemStop [0]{}%
\providecommand \bibitemNoStop [0]{.\EOS\space}%
\providecommand \EOS [0]{\spacefactor3000\relax}%
\providecommand \BibitemShut  [1]{\csname bibitem#1\endcsname}%
\let\auto@bib@innerbib\@empty
\bibitem [{\citenamefont {Bradac}\ \emph {et~al.}(2019)\citenamefont {Bradac}, \citenamefont {Gao}, \citenamefont {Forneris}, \citenamefont {Trusheim},\ and\ \citenamefont {Aharonovich}}]{Bradac2019}%
  \BibitemOpen
  \bibfield  {author} {\bibinfo {author} {\bibfnamefont {C.}~\bibnamefont {Bradac}}, \bibinfo {author} {\bibfnamefont {W.}~\bibnamefont {Gao}}, \bibinfo {author} {\bibfnamefont {J.}~\bibnamefont {Forneris}}, \bibinfo {author} {\bibfnamefont {M.~E.}\ \bibnamefont {Trusheim}}, \ and\ \bibinfo {author} {\bibfnamefont {I.}~\bibnamefont {Aharonovich}},\ }\bibfield  {title} {\enquote {\bibinfo {title} {Quantum nanophotonics with group iv defects in diamond},}\ }\href {\doibase 10.1038/s41467-019-13332-w} {\bibfield  {journal} {\bibinfo  {journal} {Nature Communications}\ }\textbf {\bibinfo {volume} {10}},\ \bibinfo {pages} {5625} (\bibinfo {year} {2019})}\BibitemShut {NoStop}%
\bibitem [{\citenamefont {Ruf}\ \emph {et~al.}(2021)\citenamefont {Ruf}, \citenamefont {Wan}, \citenamefont {Choi}, \citenamefont {Englund},\ and\ \citenamefont {Hanson}}]{Ruf2021}%
  \BibitemOpen
  \bibfield  {author} {\bibinfo {author} {\bibfnamefont {M.}~\bibnamefont {Ruf}}, \bibinfo {author} {\bibfnamefont {N.~H.}\ \bibnamefont {Wan}}, \bibinfo {author} {\bibfnamefont {H.}~\bibnamefont {Choi}}, \bibinfo {author} {\bibfnamefont {D.}~\bibnamefont {Englund}}, \ and\ \bibinfo {author} {\bibfnamefont {R.}~\bibnamefont {Hanson}},\ }\bibfield  {title} {\enquote {\bibinfo {title} {Quantum networks based on color centers in diamond},}\ }\href {\doibase 10.1063/5.0056534} {\bibfield  {journal} {\bibinfo  {journal} {Journal of Applied Physics}\ }\textbf {\bibinfo {volume} {130}},\ \bibinfo {pages} {070901} (\bibinfo {year} {2021})}\BibitemShut {NoStop}%
\bibitem [{\citenamefont {Castelletto}\ and\ \citenamefont {Boretti}(2020)}]{Castelletto2020}%
  \BibitemOpen
  \bibfield  {author} {\bibinfo {author} {\bibfnamefont {S.}~\bibnamefont {Castelletto}}\ and\ \bibinfo {author} {\bibfnamefont {A.}~\bibnamefont {Boretti}},\ }\href {\doibase 10.1088/2515-7647/ab77a2} {\enquote {\bibinfo {title} {Silicon carbide color centers for quantum applications},}\ } (\bibinfo {year} {2020})\BibitemShut {NoStop}%
\bibitem [{\citenamefont {Tiurev}\ \emph {et~al.}(2022)\citenamefont {Tiurev}, \citenamefont {Appel}, \citenamefont {Mirambell}, \citenamefont {Lauritzen}, \citenamefont {Tiranov}, \citenamefont {Lodahl},\ and\ \citenamefont {Sørensen}}]{Tiurev2022}%
  \BibitemOpen
  \bibfield  {author} {\bibinfo {author} {\bibfnamefont {K.}~\bibnamefont {Tiurev}}, \bibinfo {author} {\bibfnamefont {M.~H.}\ \bibnamefont {Appel}}, \bibinfo {author} {\bibfnamefont {P.~L.}\ \bibnamefont {Mirambell}}, \bibinfo {author} {\bibfnamefont {M.~B.}\ \bibnamefont {Lauritzen}}, \bibinfo {author} {\bibfnamefont {A.}~\bibnamefont {Tiranov}}, \bibinfo {author} {\bibfnamefont {P.}~\bibnamefont {Lodahl}}, \ and\ \bibinfo {author} {\bibfnamefont {A.~S.}\ \bibnamefont {Sørensen}},\ }\bibfield  {title} {\enquote {\bibinfo {title} {High-fidelity multiphoton-entangled cluster state with solid-state quantum emitters in photonic nanostructures},}\ }\href {\doibase 10.1103/PhysRevA.105.L030601} {\bibfield  {journal} {\bibinfo  {journal} {Physical Review A}\ }\textbf {\bibinfo {volume} {105}} (\bibinfo {year} {2022}),\ 10.1103/PhysRevA.105.L030601}\BibitemShut {NoStop}%
\bibitem [{\citenamefont {Kimble}(2008)}]{Kimble2008}%
  \BibitemOpen
  \bibfield  {author} {\bibinfo {author} {\bibfnamefont {H.~J.}\ \bibnamefont {Kimble}},\ }\href {\doibase 10.1038/nature07127} {\enquote {\bibinfo {title} {The quantum internet},}\ } (\bibinfo {year} {2008})\BibitemShut {NoStop}%
\bibitem [{\citenamefont {Evans}\ \emph {et~al.}(2016)\citenamefont {Evans}, \citenamefont {Sipahigil}, \citenamefont {Sukachev}, \citenamefont {Zibrov},\ and\ \citenamefont {Lukin}}]{Evans2016}%
  \BibitemOpen
  \bibfield  {author} {\bibinfo {author} {\bibfnamefont {R.~E.}\ \bibnamefont {Evans}}, \bibinfo {author} {\bibfnamefont {A.}~\bibnamefont {Sipahigil}}, \bibinfo {author} {\bibfnamefont {D.~D.}\ \bibnamefont {Sukachev}}, \bibinfo {author} {\bibfnamefont {A.~S.}\ \bibnamefont {Zibrov}}, \ and\ \bibinfo {author} {\bibfnamefont {M.~D.}\ \bibnamefont {Lukin}},\ }\bibfield  {title} {\enquote {\bibinfo {title} {Narrow-linewidth homogeneous optical emitters in diamond nanostructures via silicon ion implantation},}\ }\href {\doibase 10.1103/PhysRevApplied.5.044010} {\bibfield  {journal} {\bibinfo  {journal} {Physical Review Applied}\ }\textbf {\bibinfo {volume} {5}},\ \bibinfo {pages} {044010} (\bibinfo {year} {2016})}\BibitemShut {NoStop}%
\bibitem [{\citenamefont {Nguyen}\ \emph {et~al.}(2019)\citenamefont {Nguyen}, \citenamefont {Sukachev}, \citenamefont {Bhaskar}, \citenamefont {Machielse}, \citenamefont {Levonian}, \citenamefont {Knall}, \citenamefont {Stroganov}, \citenamefont {Chia}, \citenamefont {Burek}, \citenamefont {Riedinger}, \citenamefont {Park}, \citenamefont {Lončar},\ and\ \citenamefont {Lukin}}]{Nguyen2019}%
  \BibitemOpen
  \bibfield  {author} {\bibinfo {author} {\bibfnamefont {C.~T.}\ \bibnamefont {Nguyen}}, \bibinfo {author} {\bibfnamefont {D.~D.}\ \bibnamefont {Sukachev}}, \bibinfo {author} {\bibfnamefont {M.~K.}\ \bibnamefont {Bhaskar}}, \bibinfo {author} {\bibfnamefont {B.}~\bibnamefont {Machielse}}, \bibinfo {author} {\bibfnamefont {D.~S.}\ \bibnamefont {Levonian}}, \bibinfo {author} {\bibfnamefont {E.~N.}\ \bibnamefont {Knall}}, \bibinfo {author} {\bibfnamefont {P.}~\bibnamefont {Stroganov}}, \bibinfo {author} {\bibfnamefont {C.}~\bibnamefont {Chia}}, \bibinfo {author} {\bibfnamefont {M.~J.}\ \bibnamefont {Burek}}, \bibinfo {author} {\bibfnamefont {R.}~\bibnamefont {Riedinger}}, \bibinfo {author} {\bibfnamefont {H.}~\bibnamefont {Park}}, \bibinfo {author} {\bibfnamefont {M.}~\bibnamefont {Lončar}}, \ and\ \bibinfo {author} {\bibfnamefont {M.~D.}\ \bibnamefont {Lukin}},\ }\bibfield  {title} {\enquote {\bibinfo {title} {An integrated nanophotonic quantum register based on silicon-vacancy spins in diamond},}\ }\href
  {\doibase 10.1103/PhysRevB.100.165428} {\bibfield  {journal} {\bibinfo  {journal} {Physical Review B}\ }\textbf {\bibinfo {volume} {100}},\ \bibinfo {pages} {165428} (\bibinfo {year} {2019})}\BibitemShut {NoStop}%
\bibitem [{\citenamefont {Knall}\ \emph {et~al.}(2022)\citenamefont {Knall}, \citenamefont {Knaut}, \citenamefont {Bekenstein}, \citenamefont {Assumpcao}, \citenamefont {Stroganov}, \citenamefont {Gong}, \citenamefont {Huan}, \citenamefont {Stas}, \citenamefont {Machielse}, \citenamefont {Chalupnik}, \citenamefont {Levonian}, \citenamefont {Suleymanzade}, \citenamefont {Riedinger}, \citenamefont {Park}, \citenamefont {Lončar}, \citenamefont {Bhaskar},\ and\ \citenamefont {Lukin}}]{Knall2022}%
  \BibitemOpen
  \bibfield  {author} {\bibinfo {author} {\bibfnamefont {E.~N.}\ \bibnamefont {Knall}}, \bibinfo {author} {\bibfnamefont {C.~M.}\ \bibnamefont {Knaut}}, \bibinfo {author} {\bibfnamefont {R.}~\bibnamefont {Bekenstein}}, \bibinfo {author} {\bibfnamefont {D.~R.}\ \bibnamefont {Assumpcao}}, \bibinfo {author} {\bibfnamefont {P.~L.}\ \bibnamefont {Stroganov}}, \bibinfo {author} {\bibfnamefont {W.}~\bibnamefont {Gong}}, \bibinfo {author} {\bibfnamefont {Y.~Q.}\ \bibnamefont {Huan}}, \bibinfo {author} {\bibfnamefont {P.~J.}\ \bibnamefont {Stas}}, \bibinfo {author} {\bibfnamefont {B.}~\bibnamefont {Machielse}}, \bibinfo {author} {\bibfnamefont {M.}~\bibnamefont {Chalupnik}}, \bibinfo {author} {\bibfnamefont {D.}~\bibnamefont {Levonian}}, \bibinfo {author} {\bibfnamefont {A.}~\bibnamefont {Suleymanzade}}, \bibinfo {author} {\bibfnamefont {R.}~\bibnamefont {Riedinger}}, \bibinfo {author} {\bibfnamefont {H.}~\bibnamefont {Park}}, \bibinfo {author} {\bibfnamefont {M.}~\bibnamefont {Lončar}}, \bibinfo {author}
  {\bibfnamefont {M.~K.}\ \bibnamefont {Bhaskar}}, \ and\ \bibinfo {author} {\bibfnamefont {M.~D.}\ \bibnamefont {Lukin}},\ }\bibfield  {title} {\enquote {\bibinfo {title} {Efficient source of shaped single photons based on an integrated diamond nanophotonic system},}\ }\href {\doibase 10.1103/PhysRevLett.129.053603} {\bibfield  {journal} {\bibinfo  {journal} {Physical Review Letters}\ }\textbf {\bibinfo {volume} {129}} (\bibinfo {year} {2022}),\ 10.1103/PhysRevLett.129.053603}\BibitemShut {NoStop}%
\bibitem [{\citenamefont {Bhaskar}\ \emph {et~al.}(2020)\citenamefont {Bhaskar}, \citenamefont {Riedinger}, \citenamefont {Machielse}, \citenamefont {Levonian}, \citenamefont {Nguyen}, \citenamefont {Knall}, \citenamefont {Park}, \citenamefont {Englund}, \citenamefont {Lončar}, \citenamefont {Sukachev},\ and\ \citenamefont {Lukin}}]{Bhaskar2020}%
  \BibitemOpen
  \bibfield  {author} {\bibinfo {author} {\bibfnamefont {M.~K.}\ \bibnamefont {Bhaskar}}, \bibinfo {author} {\bibfnamefont {R.}~\bibnamefont {Riedinger}}, \bibinfo {author} {\bibfnamefont {B.}~\bibnamefont {Machielse}}, \bibinfo {author} {\bibfnamefont {D.~S.}\ \bibnamefont {Levonian}}, \bibinfo {author} {\bibfnamefont {C.~T.}\ \bibnamefont {Nguyen}}, \bibinfo {author} {\bibfnamefont {E.~N.}\ \bibnamefont {Knall}}, \bibinfo {author} {\bibfnamefont {H.}~\bibnamefont {Park}}, \bibinfo {author} {\bibfnamefont {D.}~\bibnamefont {Englund}}, \bibinfo {author} {\bibfnamefont {M.}~\bibnamefont {Lončar}}, \bibinfo {author} {\bibfnamefont {D.~D.}\ \bibnamefont {Sukachev}}, \ and\ \bibinfo {author} {\bibfnamefont {M.~D.}\ \bibnamefont {Lukin}},\ }\bibfield  {title} {\enquote {\bibinfo {title} {Experimental demonstration of memory-enhanced quantum communication},}\ }\href {\doibase 10.1038/s41586-020-2103-5} {\bibfield  {journal} {\bibinfo  {journal} {Nature}\ }\textbf {\bibinfo {volume} {580}} (\bibinfo {year}
  {2020}),\ 10.1038/s41586-020-2103-5}\BibitemShut {NoStop}%
\bibitem [{\citenamefont {Meesala}\ \emph {et~al.}(2018)\citenamefont {Meesala}, \citenamefont {Sohn}, \citenamefont {Pingault}, \citenamefont {Shao}, \citenamefont {Atikian}, \citenamefont {Holzgrafe}, \citenamefont {Gündoǧan}, \citenamefont {Stavrakas}, \citenamefont {Sipahigil}, \citenamefont {Chia}, \citenamefont {Evans}, \citenamefont {Burek}, \citenamefont {Zhang}, \citenamefont {Wu}, \citenamefont {Pacheco}, \citenamefont {Abraham}, \citenamefont {Bielejec}, \citenamefont {Lukin}, \citenamefont {Atatüre},\ and\ \citenamefont {Lončar}}]{Meesala2018}%
  \BibitemOpen
  \bibfield  {author} {\bibinfo {author} {\bibfnamefont {S.}~\bibnamefont {Meesala}}, \bibinfo {author} {\bibfnamefont {Y.~I.}\ \bibnamefont {Sohn}}, \bibinfo {author} {\bibfnamefont {B.}~\bibnamefont {Pingault}}, \bibinfo {author} {\bibfnamefont {L.}~\bibnamefont {Shao}}, \bibinfo {author} {\bibfnamefont {H.~A.}\ \bibnamefont {Atikian}}, \bibinfo {author} {\bibfnamefont {J.}~\bibnamefont {Holzgrafe}}, \bibinfo {author} {\bibfnamefont {M.}~\bibnamefont {Gündoǧan}}, \bibinfo {author} {\bibfnamefont {C.}~\bibnamefont {Stavrakas}}, \bibinfo {author} {\bibfnamefont {A.}~\bibnamefont {Sipahigil}}, \bibinfo {author} {\bibfnamefont {C.}~\bibnamefont {Chia}}, \bibinfo {author} {\bibfnamefont {R.}~\bibnamefont {Evans}}, \bibinfo {author} {\bibfnamefont {M.~J.}\ \bibnamefont {Burek}}, \bibinfo {author} {\bibfnamefont {M.}~\bibnamefont {Zhang}}, \bibinfo {author} {\bibfnamefont {L.}~\bibnamefont {Wu}}, \bibinfo {author} {\bibfnamefont {J.~L.}\ \bibnamefont {Pacheco}}, \bibinfo {author} {\bibfnamefont {J.}~\bibnamefont
  {Abraham}}, \bibinfo {author} {\bibfnamefont {E.}~\bibnamefont {Bielejec}}, \bibinfo {author} {\bibfnamefont {M.~D.}\ \bibnamefont {Lukin}}, \bibinfo {author} {\bibfnamefont {M.}~\bibnamefont {Atatüre}}, \ and\ \bibinfo {author} {\bibfnamefont {M.}~\bibnamefont {Lončar}},\ }\bibfield  {title} {\enquote {\bibinfo {title} {Strain engineering of the silicon-vacancy center in diamond},}\ }\href {\doibase 10.1103/PhysRevB.97.205444} {\bibfield  {journal} {\bibinfo  {journal} {Physical Review B}\ }\textbf {\bibinfo {volume} {97}} (\bibinfo {year} {2018}),\ 10.1103/PhysRevB.97.205444}\BibitemShut {NoStop}%
\bibitem [{\citenamefont {Maity}\ \emph {et~al.}(2018)\citenamefont {Maity}, \citenamefont {Shao}, \citenamefont {Sohn}, \citenamefont {Meesala}, \citenamefont {Machielse}, \citenamefont {Bielejec}, \citenamefont {Markham},\ and\ \citenamefont {Lončar}}]{Maity2018}%
  \BibitemOpen
  \bibfield  {author} {\bibinfo {author} {\bibfnamefont {S.}~\bibnamefont {Maity}}, \bibinfo {author} {\bibfnamefont {L.}~\bibnamefont {Shao}}, \bibinfo {author} {\bibfnamefont {Y.~I.}\ \bibnamefont {Sohn}}, \bibinfo {author} {\bibfnamefont {S.}~\bibnamefont {Meesala}}, \bibinfo {author} {\bibfnamefont {B.}~\bibnamefont {Machielse}}, \bibinfo {author} {\bibfnamefont {E.}~\bibnamefont {Bielejec}}, \bibinfo {author} {\bibfnamefont {M.}~\bibnamefont {Markham}}, \ and\ \bibinfo {author} {\bibfnamefont {M.}~\bibnamefont {Lončar}},\ }\bibfield  {title} {\enquote {\bibinfo {title} {Spectral alignment of single-photon emitters in diamond using strain gradient},}\ }\href {\doibase 10.1103/PhysRevApplied.10.024050} {\bibfield  {journal} {\bibinfo  {journal} {Physical Review Applied}\ }\textbf {\bibinfo {volume} {10}} (\bibinfo {year} {2018}),\ 10.1103/PhysRevApplied.10.024050}\BibitemShut {NoStop}%
\bibitem [{\citenamefont {Falk}\ \emph {et~al.}(2014)\citenamefont {Falk}, \citenamefont {Klimov}, \citenamefont {Buckley}, \citenamefont {Ivády}, \citenamefont {Abrikosov}, \citenamefont {Calusine}, \citenamefont {Koehl}, \citenamefont {Ádám Gali},\ and\ \citenamefont {Awschalom}}]{Falk2014}%
  \BibitemOpen
  \bibfield  {author} {\bibinfo {author} {\bibfnamefont {A.~L.}\ \bibnamefont {Falk}}, \bibinfo {author} {\bibfnamefont {P.~V.}\ \bibnamefont {Klimov}}, \bibinfo {author} {\bibfnamefont {B.~B.}\ \bibnamefont {Buckley}}, \bibinfo {author} {\bibfnamefont {V.}~\bibnamefont {Ivády}}, \bibinfo {author} {\bibfnamefont {I.~A.}\ \bibnamefont {Abrikosov}}, \bibinfo {author} {\bibfnamefont {G.}~\bibnamefont {Calusine}}, \bibinfo {author} {\bibfnamefont {W.~F.}\ \bibnamefont {Koehl}}, \bibinfo {author} {\bibnamefont {Ádám Gali}}, \ and\ \bibinfo {author} {\bibfnamefont {D.~D.}\ \bibnamefont {Awschalom}},\ }\bibfield  {title} {\enquote {\bibinfo {title} {Electrically and mechanically tunable electron spins in silicon carbide color centers},}\ }\href {\doibase 10.1103/PhysRevLett.112.187601} {\bibfield  {journal} {\bibinfo  {journal} {Physical Review Letters}\ }\textbf {\bibinfo {volume} {112}} (\bibinfo {year} {2014}),\ 10.1103/PhysRevLett.112.187601}\BibitemShut {NoStop}%
\bibitem [{\citenamefont {Knauer}, \citenamefont {Hadden},\ and\ \citenamefont {Rarity}(2020)}]{Knauer2020}%
  \BibitemOpen
  \bibfield  {author} {\bibinfo {author} {\bibfnamefont {S.}~\bibnamefont {Knauer}}, \bibinfo {author} {\bibfnamefont {J.~P.}\ \bibnamefont {Hadden}}, \ and\ \bibinfo {author} {\bibfnamefont {J.~G.}\ \bibnamefont {Rarity}},\ }\bibfield  {title} {\enquote {\bibinfo {title} {In-situ measurements of fabrication induced strain in diamond photonic-structures using intrinsic colour centres},}\ }\href {\doibase 10.1038/s41534-020-0277-1} {\bibfield  {journal} {\bibinfo  {journal} {npj Quantum Information}\ }\textbf {\bibinfo {volume} {6}} (\bibinfo {year} {2020}),\ 10.1038/s41534-020-0277-1}\BibitemShut {NoStop}%
\bibitem [{\citenamefont {Batalov}\ \emph {et~al.}(2009)\citenamefont {Batalov}, \citenamefont {Jacques}, \citenamefont {Kaiser}, \citenamefont {Siyushev}, \citenamefont {Neumann}, \citenamefont {Rogers}, \citenamefont {McMurtrie}, \citenamefont {Manson}, \citenamefont {Jelezko},\ and\ \citenamefont {Wrachtrup}}]{Batalov2009}%
  \BibitemOpen
  \bibfield  {author} {\bibinfo {author} {\bibfnamefont {A.}~\bibnamefont {Batalov}}, \bibinfo {author} {\bibfnamefont {V.}~\bibnamefont {Jacques}}, \bibinfo {author} {\bibfnamefont {F.}~\bibnamefont {Kaiser}}, \bibinfo {author} {\bibfnamefont {P.}~\bibnamefont {Siyushev}}, \bibinfo {author} {\bibfnamefont {P.}~\bibnamefont {Neumann}}, \bibinfo {author} {\bibfnamefont {L.~J.}\ \bibnamefont {Rogers}}, \bibinfo {author} {\bibfnamefont {R.~L.}\ \bibnamefont {McMurtrie}}, \bibinfo {author} {\bibfnamefont {N.~B.}\ \bibnamefont {Manson}}, \bibinfo {author} {\bibfnamefont {F.}~\bibnamefont {Jelezko}}, \ and\ \bibinfo {author} {\bibfnamefont {J.}~\bibnamefont {Wrachtrup}},\ }\bibfield  {title} {\enquote {\bibinfo {title} {Low temperature studies of the excited-state structure of negatively charged nitrogen-vacancy color centers in diamond},}\ }\href {\doibase 10.1103/PhysRevLett.102.195506} {\bibfield  {journal} {\bibinfo  {journal} {Physical Review Letters}\ }\textbf {\bibinfo {volume} {102}} (\bibinfo {year}
  {2009}),\ 10.1103/PhysRevLett.102.195506}\BibitemShut {NoStop}%
\bibitem [{\citenamefont {Olivero}\ \emph {et~al.}(2013)\citenamefont {Olivero}, \citenamefont {Bosia}, \citenamefont {Fairchild}, \citenamefont {Gibson}, \citenamefont {Greentree}, \citenamefont {Spizzirri},\ and\ \citenamefont {Prawer}}]{Olivero2013}%
  \BibitemOpen
  \bibfield  {author} {\bibinfo {author} {\bibfnamefont {P.}~\bibnamefont {Olivero}}, \bibinfo {author} {\bibfnamefont {F.}~\bibnamefont {Bosia}}, \bibinfo {author} {\bibfnamefont {B.~A.}\ \bibnamefont {Fairchild}}, \bibinfo {author} {\bibfnamefont {B.~C.}\ \bibnamefont {Gibson}}, \bibinfo {author} {\bibfnamefont {A.~D.}\ \bibnamefont {Greentree}}, \bibinfo {author} {\bibfnamefont {P.}~\bibnamefont {Spizzirri}}, \ and\ \bibinfo {author} {\bibfnamefont {S.}~\bibnamefont {Prawer}},\ }\bibfield  {title} {\enquote {\bibinfo {title} {Splitting of photoluminescent emission from nitrogen-vacancy centers in diamond induced by ion-damage-induced stress},}\ }\href {\doibase 10.1088/1367-2630/15/4/043027} {\bibfield  {journal} {\bibinfo  {journal} {New Journal of Physics}\ }\textbf {\bibinfo {volume} {15}} (\bibinfo {year} {2013}),\ 10.1088/1367-2630/15/4/043027}\BibitemShut {NoStop}%
\bibitem [{\citenamefont {Udvarhelyi}\ \emph {et~al.}(2018)\citenamefont {Udvarhelyi}, \citenamefont {Shkolnikov}, \citenamefont {Gali}, \citenamefont {Burkard},\ and\ \citenamefont {Pályi}}]{Udvarhelyi2018}%
  \BibitemOpen
  \bibfield  {author} {\bibinfo {author} {\bibfnamefont {P.}~\bibnamefont {Udvarhelyi}}, \bibinfo {author} {\bibfnamefont {V.~O.}\ \bibnamefont {Shkolnikov}}, \bibinfo {author} {\bibfnamefont {A.}~\bibnamefont {Gali}}, \bibinfo {author} {\bibfnamefont {G.}~\bibnamefont {Burkard}}, \ and\ \bibinfo {author} {\bibfnamefont {A.}~\bibnamefont {Pályi}},\ }\bibfield  {title} {\enquote {\bibinfo {title} {Spin-strain interaction in nitrogen-vacancy centers in diamond},}\ }\href {\doibase 10.1103/PhysRevB.98.075201} {\bibfield  {journal} {\bibinfo  {journal} {Physical Review B}\ }\textbf {\bibinfo {volume} {98}} (\bibinfo {year} {2018}),\ 10.1103/PhysRevB.98.075201}\BibitemShut {NoStop}%
\bibitem [{\citenamefont {Li}\ \emph {et~al.}(2020)\citenamefont {Li}, \citenamefont {Chou}, \citenamefont {Hu}, \citenamefont {Plenio}, \citenamefont {Udvarhelyi}, \citenamefont {Thiering}, \citenamefont {Abdi},\ and\ \citenamefont {Gali}}]{Li2020}%
  \BibitemOpen
  \bibfield  {author} {\bibinfo {author} {\bibfnamefont {S.}~\bibnamefont {Li}}, \bibinfo {author} {\bibfnamefont {J.~P.}\ \bibnamefont {Chou}}, \bibinfo {author} {\bibfnamefont {A.}~\bibnamefont {Hu}}, \bibinfo {author} {\bibfnamefont {M.~B.}\ \bibnamefont {Plenio}}, \bibinfo {author} {\bibfnamefont {P.}~\bibnamefont {Udvarhelyi}}, \bibinfo {author} {\bibfnamefont {G.}~\bibnamefont {Thiering}}, \bibinfo {author} {\bibfnamefont {M.}~\bibnamefont {Abdi}}, \ and\ \bibinfo {author} {\bibfnamefont {A.}~\bibnamefont {Gali}},\ }\bibfield  {title} {\enquote {\bibinfo {title} {Giant shift upon strain on the fluorescence spectrum of vnnb color centers in h-bn},}\ }\href {\doibase 10.1038/s41534-020-00312-y} {\bibfield  {journal} {\bibinfo  {journal} {npj Quantum Information}\ }\textbf {\bibinfo {volume} {6}} (\bibinfo {year} {2020}),\ 10.1038/s41534-020-00312-y}\BibitemShut {NoStop}%
\bibitem [{\citenamefont {Lindner}\ \emph {et~al.}(2018)\citenamefont {Lindner}, \citenamefont {Bommer}, \citenamefont {Muzha}, \citenamefont {Krueger}, \citenamefont {Gines}, \citenamefont {Mandal}, \citenamefont {Williams}, \citenamefont {Londero}, \citenamefont {Gali},\ and\ \citenamefont {Becher}}]{Lindner2018}%
  \BibitemOpen
  \bibfield  {author} {\bibinfo {author} {\bibfnamefont {S.}~\bibnamefont {Lindner}}, \bibinfo {author} {\bibfnamefont {A.}~\bibnamefont {Bommer}}, \bibinfo {author} {\bibfnamefont {A.}~\bibnamefont {Muzha}}, \bibinfo {author} {\bibfnamefont {A.}~\bibnamefont {Krueger}}, \bibinfo {author} {\bibfnamefont {L.}~\bibnamefont {Gines}}, \bibinfo {author} {\bibfnamefont {S.}~\bibnamefont {Mandal}}, \bibinfo {author} {\bibfnamefont {O.}~\bibnamefont {Williams}}, \bibinfo {author} {\bibfnamefont {E.}~\bibnamefont {Londero}}, \bibinfo {author} {\bibfnamefont {A.}~\bibnamefont {Gali}}, \ and\ \bibinfo {author} {\bibfnamefont {C.}~\bibnamefont {Becher}},\ }\bibfield  {title} {\enquote {\bibinfo {title} {Strongly inhomogeneous distribution of spectral properties of silicon-vacancy color centers in nanodiamonds},}\ }\href {\doibase 10.1088/1367-2630/aae93f} {\bibfield  {journal} {\bibinfo  {journal} {New Journal of Physics}\ }\textbf {\bibinfo {volume} {20}} (\bibinfo {year} {2018}),\ 10.1088/1367-2630/aae93f}\BibitemShut
  {NoStop}%
\bibitem [{\citenamefont {Sukachev}\ \emph {et~al.}(2017)\citenamefont {Sukachev}, \citenamefont {Sipahigil}, \citenamefont {Nguyen}, \citenamefont {Bhaskar}, \citenamefont {Evans}, \citenamefont {Jelezko},\ and\ \citenamefont {Lukin}}]{Sukachev2017}%
  \BibitemOpen
  \bibfield  {author} {\bibinfo {author} {\bibfnamefont {D.~D.}\ \bibnamefont {Sukachev}}, \bibinfo {author} {\bibfnamefont {A.}~\bibnamefont {Sipahigil}}, \bibinfo {author} {\bibfnamefont {C.~T.}\ \bibnamefont {Nguyen}}, \bibinfo {author} {\bibfnamefont {M.~K.}\ \bibnamefont {Bhaskar}}, \bibinfo {author} {\bibfnamefont {R.~E.}\ \bibnamefont {Evans}}, \bibinfo {author} {\bibfnamefont {F.}~\bibnamefont {Jelezko}}, \ and\ \bibinfo {author} {\bibfnamefont {M.~D.}\ \bibnamefont {Lukin}},\ }\bibfield  {title} {\enquote {\bibinfo {title} {Silicon-vacancy spin qubit in diamond: A quantum memory exceeding 10 ms with single-shot state readout},}\ }\href {\doibase 10.1103/PhysRevLett.119.223602} {\bibfield  {journal} {\bibinfo  {journal} {Physical Review Letters}\ }\textbf {\bibinfo {volume} {119}} (\bibinfo {year} {2017}),\ 10.1103/PhysRevLett.119.223602}\BibitemShut {NoStop}%
\bibitem [{\citenamefont {Sohn}\ \emph {et~al.}(2018)\citenamefont {Sohn}, \citenamefont {Meesala}, \citenamefont {Pingault}, \citenamefont {Atikian}, \citenamefont {Holzgrafe}, \citenamefont {Gündoǧan}, \citenamefont {Stavrakas}, \citenamefont {Stanley}, \citenamefont {Sipahigil}, \citenamefont {Choi}, \citenamefont {Zhang}, \citenamefont {Pacheco}, \citenamefont {Abraham}, \citenamefont {Bielejec}, \citenamefont {Lukin}, \citenamefont {Atatüre},\ and\ \citenamefont {Lončar}}]{Sohn2018}%
  \BibitemOpen
  \bibfield  {author} {\bibinfo {author} {\bibfnamefont {Y.~I.}\ \bibnamefont {Sohn}}, \bibinfo {author} {\bibfnamefont {S.}~\bibnamefont {Meesala}}, \bibinfo {author} {\bibfnamefont {B.}~\bibnamefont {Pingault}}, \bibinfo {author} {\bibfnamefont {H.~A.}\ \bibnamefont {Atikian}}, \bibinfo {author} {\bibfnamefont {J.}~\bibnamefont {Holzgrafe}}, \bibinfo {author} {\bibfnamefont {M.}~\bibnamefont {Gündoǧan}}, \bibinfo {author} {\bibfnamefont {C.}~\bibnamefont {Stavrakas}}, \bibinfo {author} {\bibfnamefont {M.~J.}\ \bibnamefont {Stanley}}, \bibinfo {author} {\bibfnamefont {A.}~\bibnamefont {Sipahigil}}, \bibinfo {author} {\bibfnamefont {J.}~\bibnamefont {Choi}}, \bibinfo {author} {\bibfnamefont {M.}~\bibnamefont {Zhang}}, \bibinfo {author} {\bibfnamefont {J.~L.}\ \bibnamefont {Pacheco}}, \bibinfo {author} {\bibfnamefont {J.}~\bibnamefont {Abraham}}, \bibinfo {author} {\bibfnamefont {E.}~\bibnamefont {Bielejec}}, \bibinfo {author} {\bibfnamefont {M.~D.}\ \bibnamefont {Lukin}}, \bibinfo {author} {\bibfnamefont
  {M.}~\bibnamefont {Atatüre}}, \ and\ \bibinfo {author} {\bibfnamefont {M.}~\bibnamefont {Lončar}},\ }\bibfield  {title} {\enquote {\bibinfo {title} {Controlling the coherence of a diamond spin qubit through its strain environment},}\ }\href {\doibase 10.1038/s41467-018-04340-3} {\bibfield  {journal} {\bibinfo  {journal} {Nature Communications}\ }\textbf {\bibinfo {volume} {9}} (\bibinfo {year} {2018}),\ 10.1038/s41467-018-04340-3}\BibitemShut {NoStop}%
\bibitem [{\citenamefont {Stas}\ \emph {et~al.}(2022)\citenamefont {Stas}, \citenamefont {Huan}, \citenamefont {Machielse}, \citenamefont {Knall}, \citenamefont {Suleymanzade}, \citenamefont {Pingault}, \citenamefont {Sutula}, \citenamefont {Ding}, \citenamefont {Knaut}, \citenamefont {Assumpcao}, \citenamefont {Wei}, \citenamefont {Bhaskar}, \citenamefont {Riedinger}, \citenamefont {Sukachev}, \citenamefont {Park}, \citenamefont {Lončar}, \citenamefont {Levonian},\ and\ \citenamefont {Lukin}}]{Stas2022}%
  \BibitemOpen
  \bibfield  {author} {\bibinfo {author} {\bibfnamefont {P.~J.}\ \bibnamefont {Stas}}, \bibinfo {author} {\bibfnamefont {Y.~Q.}\ \bibnamefont {Huan}}, \bibinfo {author} {\bibfnamefont {B.}~\bibnamefont {Machielse}}, \bibinfo {author} {\bibfnamefont {E.~N.}\ \bibnamefont {Knall}}, \bibinfo {author} {\bibfnamefont {A.}~\bibnamefont {Suleymanzade}}, \bibinfo {author} {\bibfnamefont {B.}~\bibnamefont {Pingault}}, \bibinfo {author} {\bibfnamefont {M.}~\bibnamefont {Sutula}}, \bibinfo {author} {\bibfnamefont {S.~W.}\ \bibnamefont {Ding}}, \bibinfo {author} {\bibfnamefont {C.~M.}\ \bibnamefont {Knaut}}, \bibinfo {author} {\bibfnamefont {D.~R.}\ \bibnamefont {Assumpcao}}, \bibinfo {author} {\bibfnamefont {Y.~C.}\ \bibnamefont {Wei}}, \bibinfo {author} {\bibfnamefont {M.~K.}\ \bibnamefont {Bhaskar}}, \bibinfo {author} {\bibfnamefont {R.}~\bibnamefont {Riedinger}}, \bibinfo {author} {\bibfnamefont {D.~D.}\ \bibnamefont {Sukachev}}, \bibinfo {author} {\bibfnamefont {H.}~\bibnamefont {Park}}, \bibinfo {author}
  {\bibfnamefont {M.}~\bibnamefont {Lončar}}, \bibinfo {author} {\bibfnamefont {D.~S.}\ \bibnamefont {Levonian}}, \ and\ \bibinfo {author} {\bibfnamefont {M.~D.}\ \bibnamefont {Lukin}},\ }\bibfield  {title} {\enquote {\bibinfo {title} {Robust multi-qubit quantum network node with integrated error detection},}\ }\href {\doibase 10.1126/science.add9771} {\bibfield  {journal} {\bibinfo  {journal} {Science}\ }\textbf {\bibinfo {volume} {378}} (\bibinfo {year} {2022}),\ 10.1126/science.add9771}\BibitemShut {NoStop}%
\bibitem [{\citenamefont {Machielse}\ \emph {et~al.}(2019)\citenamefont {Machielse}, \citenamefont {Bogdanovic}, \citenamefont {Meesala}, \citenamefont {Gauthier}, \citenamefont {Burek}, \citenamefont {Joe}, \citenamefont {Chalupnik}, \citenamefont {Sohn}, \citenamefont {Holzgrafe}, \citenamefont {Evans}, \citenamefont {Chia}, \citenamefont {Atikian}, \citenamefont {Bhaskar}, \citenamefont {Sukachev}, \citenamefont {Shao}, \citenamefont {Maity}, \citenamefont {Lukin},\ and\ \citenamefont {Lončar}}]{Machielse2019}%
  \BibitemOpen
  \bibfield  {author} {\bibinfo {author} {\bibfnamefont {B.}~\bibnamefont {Machielse}}, \bibinfo {author} {\bibfnamefont {S.}~\bibnamefont {Bogdanovic}}, \bibinfo {author} {\bibfnamefont {S.}~\bibnamefont {Meesala}}, \bibinfo {author} {\bibfnamefont {S.}~\bibnamefont {Gauthier}}, \bibinfo {author} {\bibfnamefont {M.}~\bibnamefont {Burek}}, \bibinfo {author} {\bibfnamefont {G.}~\bibnamefont {Joe}}, \bibinfo {author} {\bibfnamefont {M.}~\bibnamefont {Chalupnik}}, \bibinfo {author} {\bibfnamefont {Y.}~\bibnamefont {Sohn}}, \bibinfo {author} {\bibfnamefont {J.}~\bibnamefont {Holzgrafe}}, \bibinfo {author} {\bibfnamefont {R.}~\bibnamefont {Evans}}, \bibinfo {author} {\bibfnamefont {C.}~\bibnamefont {Chia}}, \bibinfo {author} {\bibfnamefont {H.}~\bibnamefont {Atikian}}, \bibinfo {author} {\bibfnamefont {M.}~\bibnamefont {Bhaskar}}, \bibinfo {author} {\bibfnamefont {D.}~\bibnamefont {Sukachev}}, \bibinfo {author} {\bibfnamefont {L.}~\bibnamefont {Shao}}, \bibinfo {author} {\bibfnamefont {S.}~\bibnamefont {Maity}},
  \bibinfo {author} {\bibfnamefont {M.}~\bibnamefont {Lukin}}, \ and\ \bibinfo {author} {\bibfnamefont {M.}~\bibnamefont {Lončar}},\ }\bibfield  {title} {\enquote {\bibinfo {title} {Quantum interference of electromechanically stabilized emitters in nanophotonic devices},}\ }\href {\doibase 10.1103/PhysRevX.9.031022} {\bibfield  {journal} {\bibinfo  {journal} {Physical Review X}\ }\textbf {\bibinfo {volume} {9}},\ \bibinfo {pages} {031022} (\bibinfo {year} {2019})}\BibitemShut {NoStop}%
\bibitem [{\citenamefont {Guo}\ \emph {et~al.}(2023)\citenamefont {Guo}, \citenamefont {Stramma}, \citenamefont {Li}, \citenamefont {Roth}, \citenamefont {Huang}, \citenamefont {Jin}, \citenamefont {Parker}, \citenamefont {Martínez}, \citenamefont {Shofer}, \citenamefont {Michaels}, \citenamefont {Purser}, \citenamefont {Appel}, \citenamefont {Alexeev}, \citenamefont {Liu}, \citenamefont {Ferrari}, \citenamefont {Awschalom}, \citenamefont {Delegan}, \citenamefont {Pingault}, \citenamefont {Galli}, \citenamefont {Heremans}, \citenamefont {Atatüre},\ and\ \citenamefont {High}}]{guo2023microwavebased}%
  \BibitemOpen
  \bibfield  {author} {\bibinfo {author} {\bibfnamefont {X.}~\bibnamefont {Guo}}, \bibinfo {author} {\bibfnamefont {A.~M.}\ \bibnamefont {Stramma}}, \bibinfo {author} {\bibfnamefont {Z.}~\bibnamefont {Li}}, \bibinfo {author} {\bibfnamefont {W.~G.}\ \bibnamefont {Roth}}, \bibinfo {author} {\bibfnamefont {B.}~\bibnamefont {Huang}}, \bibinfo {author} {\bibfnamefont {Y.}~\bibnamefont {Jin}}, \bibinfo {author} {\bibfnamefont {R.~A.}\ \bibnamefont {Parker}}, \bibinfo {author} {\bibfnamefont {J.~A.}\ \bibnamefont {Martínez}}, \bibinfo {author} {\bibfnamefont {N.}~\bibnamefont {Shofer}}, \bibinfo {author} {\bibfnamefont {C.~P.}\ \bibnamefont {Michaels}}, \bibinfo {author} {\bibfnamefont {C.~P.}\ \bibnamefont {Purser}}, \bibinfo {author} {\bibfnamefont {M.~H.}\ \bibnamefont {Appel}}, \bibinfo {author} {\bibfnamefont {E.~M.}\ \bibnamefont {Alexeev}}, \bibinfo {author} {\bibfnamefont {T.}~\bibnamefont {Liu}}, \bibinfo {author} {\bibfnamefont {A.~C.}\ \bibnamefont {Ferrari}}, \bibinfo {author} {\bibfnamefont {D.~D.}\
  \bibnamefont {Awschalom}}, \bibinfo {author} {\bibfnamefont {N.}~\bibnamefont {Delegan}}, \bibinfo {author} {\bibfnamefont {B.}~\bibnamefont {Pingault}}, \bibinfo {author} {\bibfnamefont {G.}~\bibnamefont {Galli}}, \bibinfo {author} {\bibfnamefont {F.~J.}\ \bibnamefont {Heremans}}, \bibinfo {author} {\bibfnamefont {M.}~\bibnamefont {Atatüre}}, \ and\ \bibinfo {author} {\bibfnamefont {A.~A.}\ \bibnamefont {High}},\ }\href@noop {} {\enquote {\bibinfo {title} {Microwave-based quantum control and coherence protection of tin-vacancy spin qubits in a strain-tuned diamond membrane heterostructure},}\ } (\bibinfo {year} {2023}),\ \Eprint {http://arxiv.org/abs/2307.11916} {arXiv:2307.11916 [cond-mat.mes-hall]} \BibitemShut {NoStop}%
\bibitem [{\citenamefont {Chia}\ \emph {et~al.}(2022)\citenamefont {Chia}, \citenamefont {Machielse}, \citenamefont {Shams-Ansari},\ and\ \citenamefont {Lončar}}]{Chia2022}%
  \BibitemOpen
  \bibfield  {author} {\bibinfo {author} {\bibfnamefont {C.}~\bibnamefont {Chia}}, \bibinfo {author} {\bibfnamefont {B.}~\bibnamefont {Machielse}}, \bibinfo {author} {\bibfnamefont {A.}~\bibnamefont {Shams-Ansari}}, \ and\ \bibinfo {author} {\bibfnamefont {M.}~\bibnamefont {Lončar}},\ }\bibfield  {title} {\enquote {\bibinfo {title} {Development of hard masks for reactive ion beam angled etching of diamond},}\ }\href {\doibase 10.1364/OE.452826} {\bibfield  {journal} {\bibinfo  {journal} {Optics Express}\ }\textbf {\bibinfo {volume} {30}},\ \bibinfo {pages} {14189} (\bibinfo {year} {2022})}\BibitemShut {NoStop}%
\bibitem [{\citenamefont {Jacobsen}\ \emph {et~al.}(2006)\citenamefont {Jacobsen}, \citenamefont {Andersen}, \citenamefont {Borel}, \citenamefont {Fage-Pedersen}, \citenamefont {Frandsen}, \citenamefont {Hansen}, \citenamefont {Kristensen}, \citenamefont {Lavrinenko}, \citenamefont {Moulin}, \citenamefont {Ou}, \citenamefont {Peucheret}, \citenamefont {Zsigri},\ and\ \citenamefont {Bjarklev}}]{Jacobsen2006}%
  \BibitemOpen
  \bibfield  {author} {\bibinfo {author} {\bibfnamefont {R.~S.}\ \bibnamefont {Jacobsen}}, \bibinfo {author} {\bibfnamefont {K.~N.}\ \bibnamefont {Andersen}}, \bibinfo {author} {\bibfnamefont {P.~I.}\ \bibnamefont {Borel}}, \bibinfo {author} {\bibfnamefont {J.}~\bibnamefont {Fage-Pedersen}}, \bibinfo {author} {\bibfnamefont {L.~H.}\ \bibnamefont {Frandsen}}, \bibinfo {author} {\bibfnamefont {O.}~\bibnamefont {Hansen}}, \bibinfo {author} {\bibfnamefont {M.}~\bibnamefont {Kristensen}}, \bibinfo {author} {\bibfnamefont {A.~V.}\ \bibnamefont {Lavrinenko}}, \bibinfo {author} {\bibfnamefont {G.}~\bibnamefont {Moulin}}, \bibinfo {author} {\bibfnamefont {H.}~\bibnamefont {Ou}}, \bibinfo {author} {\bibfnamefont {C.}~\bibnamefont {Peucheret}}, \bibinfo {author} {\bibfnamefont {B.}~\bibnamefont {Zsigri}}, \ and\ \bibinfo {author} {\bibfnamefont {A.}~\bibnamefont {Bjarklev}},\ }\bibfield  {title} {\enquote {\bibinfo {title} {Strained silicon as a new electro-optic material},}\ }\href {\doibase 10.1038/nature04706}
  {\bibfield  {journal} {\bibinfo  {journal} {Nature}\ }\textbf {\bibinfo {volume} {441}} (\bibinfo {year} {2006}),\ 10.1038/nature04706}\BibitemShut {NoStop}%
\bibitem [{\citenamefont {Ke}, \citenamefont {Chrostowski},\ and\ \citenamefont {Xia}(2017)}]{Ke2017}%
  \BibitemOpen
  \bibfield  {author} {\bibinfo {author} {\bibfnamefont {J.}~\bibnamefont {Ke}}, \bibinfo {author} {\bibfnamefont {L.}~\bibnamefont {Chrostowski}}, \ and\ \bibinfo {author} {\bibfnamefont {G.}~\bibnamefont {Xia}},\ }\bibfield  {title} {\enquote {\bibinfo {title} {Stress engineering with silicon nitride stressors for ge-on-si lasers},}\ }\href {\doibase 10.1109/JPHOT.2017.2675401} {\bibfield  {journal} {\bibinfo  {journal} {IEEE Photonics Journal}\ }\textbf {\bibinfo {volume} {9}} (\bibinfo {year} {2017}),\ 10.1109/JPHOT.2017.2675401}\BibitemShut {NoStop}%
\bibitem [{\citenamefont {Ghrib}\ \emph {et~al.}(2013)\citenamefont {Ghrib}, \citenamefont {Kurdi}, \citenamefont {Kersauson}, \citenamefont {Prost}, \citenamefont {Sauvage}, \citenamefont {Checoury}, \citenamefont {Beaudoin}, \citenamefont {Sagnes},\ and\ \citenamefont {Boucaud}}]{Ghrib2013}%
  \BibitemOpen
  \bibfield  {author} {\bibinfo {author} {\bibfnamefont {A.}~\bibnamefont {Ghrib}}, \bibinfo {author} {\bibfnamefont {M.~E.}\ \bibnamefont {Kurdi}}, \bibinfo {author} {\bibfnamefont {M.~D.}\ \bibnamefont {Kersauson}}, \bibinfo {author} {\bibfnamefont {M.}~\bibnamefont {Prost}}, \bibinfo {author} {\bibfnamefont {S.}~\bibnamefont {Sauvage}}, \bibinfo {author} {\bibfnamefont {X.}~\bibnamefont {Checoury}}, \bibinfo {author} {\bibfnamefont {G.}~\bibnamefont {Beaudoin}}, \bibinfo {author} {\bibfnamefont {I.}~\bibnamefont {Sagnes}}, \ and\ \bibinfo {author} {\bibfnamefont {P.}~\bibnamefont {Boucaud}},\ }\bibfield  {title} {\enquote {\bibinfo {title} {Tensile-strained germanium microdisks},}\ }\href {\doibase 10.1063/1.4809832} {\bibfield  {journal} {\bibinfo  {journal} {Applied Physics Letters}\ }\textbf {\bibinfo {volume} {102}} (\bibinfo {year} {2013}),\ 10.1063/1.4809832}\BibitemShut {NoStop}%
\bibitem [{\citenamefont {Sukhdeo}\ \emph {et~al.}(2015)\citenamefont {Sukhdeo}, \citenamefont {Petykiewicz}, \citenamefont {Gupta}, \citenamefont {Kim}, \citenamefont {Woo}, \citenamefont {Kim}, \citenamefont {Vučković}, \citenamefont {Saraswat},\ and\ \citenamefont {Nam}}]{Sukhdeo2015}%
  \BibitemOpen
  \bibfield  {author} {\bibinfo {author} {\bibfnamefont {D.~S.}\ \bibnamefont {Sukhdeo}}, \bibinfo {author} {\bibfnamefont {J.}~\bibnamefont {Petykiewicz}}, \bibinfo {author} {\bibfnamefont {S.}~\bibnamefont {Gupta}}, \bibinfo {author} {\bibfnamefont {D.}~\bibnamefont {Kim}}, \bibinfo {author} {\bibfnamefont {S.}~\bibnamefont {Woo}}, \bibinfo {author} {\bibfnamefont {Y.}~\bibnamefont {Kim}}, \bibinfo {author} {\bibfnamefont {J.}~\bibnamefont {Vučković}}, \bibinfo {author} {\bibfnamefont {K.~C.}\ \bibnamefont {Saraswat}}, \ and\ \bibinfo {author} {\bibfnamefont {D.}~\bibnamefont {Nam}},\ }\bibfield  {title} {\enquote {\bibinfo {title} {Ge microdisk with lithographically-tunable strain using cmos-compatible process},}\ }\href {\doibase 10.1364/oe.23.033249} {\bibfield  {journal} {\bibinfo  {journal} {Optics Express}\ }\textbf {\bibinfo {volume} {23}} (\bibinfo {year} {2015}),\ 10.1364/oe.23.033249}\BibitemShut {NoStop}%
\bibitem [{\citenamefont {Balseanu}\ \emph {et~al.}(2006)\citenamefont {Balseanu}, \citenamefont {Xia}, \citenamefont {Zubkov}, \citenamefont {Le}, \citenamefont {Lee},\ and\ \citenamefont {M'Saad}}]{Balseanu2006}%
  \BibitemOpen
  \bibfield  {author} {\bibinfo {author} {\bibfnamefont {M.}~\bibnamefont {Balseanu}}, \bibinfo {author} {\bibfnamefont {L.-Q.}\ \bibnamefont {Xia}}, \bibinfo {author} {\bibfnamefont {V.}~\bibnamefont {Zubkov}}, \bibinfo {author} {\bibfnamefont {M.}~\bibnamefont {Le}}, \bibinfo {author} {\bibfnamefont {J.}~\bibnamefont {Lee}}, \ and\ \bibinfo {author} {\bibfnamefont {H.}~\bibnamefont {M'Saad}},\ }\bibfield  {title} {\enquote {\bibinfo {title} {Stress modulation of pecvd silicon nitride},}\ }\href {\doibase 10.1149/ma2005-02/13/532} {\bibfield  {journal} {\bibinfo  {journal} {ECS Meeting Abstracts}\ }\textbf {\bibinfo {volume} {MA2005-02}} (\bibinfo {year} {2006}),\ 10.1149/ma2005-02/13/532}\BibitemShut {NoStop}%
\bibitem [{\citenamefont {Grim}\ \emph {et~al.}(2019)\citenamefont {Grim}, \citenamefont {Bracker}, \citenamefont {Zalalutdinov}, \citenamefont {Carter}, \citenamefont {Kozen}, \citenamefont {Kim}, \citenamefont {Kim}, \citenamefont {Mlack}, \citenamefont {Yakes}, \citenamefont {Lee},\ and\ \citenamefont {Gammon}}]{Grim2019}%
  \BibitemOpen
  \bibfield  {author} {\bibinfo {author} {\bibfnamefont {J.~Q.}\ \bibnamefont {Grim}}, \bibinfo {author} {\bibfnamefont {A.~S.}\ \bibnamefont {Bracker}}, \bibinfo {author} {\bibfnamefont {M.}~\bibnamefont {Zalalutdinov}}, \bibinfo {author} {\bibfnamefont {S.~G.}\ \bibnamefont {Carter}}, \bibinfo {author} {\bibfnamefont {A.~C.}\ \bibnamefont {Kozen}}, \bibinfo {author} {\bibfnamefont {M.}~\bibnamefont {Kim}}, \bibinfo {author} {\bibfnamefont {C.~S.}\ \bibnamefont {Kim}}, \bibinfo {author} {\bibfnamefont {J.~T.}\ \bibnamefont {Mlack}}, \bibinfo {author} {\bibfnamefont {M.}~\bibnamefont {Yakes}}, \bibinfo {author} {\bibfnamefont {B.}~\bibnamefont {Lee}}, \ and\ \bibinfo {author} {\bibfnamefont {D.}~\bibnamefont {Gammon}},\ }\bibfield  {title} {\enquote {\bibinfo {title} {Scalable in operando strain tuning in nanophotonic waveguides enabling three-quantum-dot superradiance},}\ }\href {\doibase 10.1038/s41563-019-0418-0} {\bibfield  {journal} {\bibinfo  {journal} {Nature Materials}\ }\textbf {\bibinfo {volume}
  {18}} (\bibinfo {year} {2019}),\ 10.1038/s41563-019-0418-0}\BibitemShut {NoStop}%
\end{thebibliography}%

\end{document}



\title[]{Deterministic Creation of Strained Color Centers in Nanostructures via High-Stress Thin Films: Supporting Information}
\author{D. R.  Assumpcao}
\author{C. Jin}%
\affiliation{ 
John A. Paulson School of Engineering and Applied Sciences, Harvard University, Cambridge, MA, 02138, USA
}%

\author{M. Sutula}

\affiliation{ 
Department of Physics , Harvard University, Cambridge, MA, 02138, USA
}%

\author{S. W. Ding}
\affiliation{ 
John A. Paulson School of Engineering and Applied Sciences, Harvard University, Cambridge, MA, 02138, USA
}%

\author{P. Pham}
\affiliation{ 
John A. Paulson School of Engineering and Applied Sciences, Harvard University, Cambridge, MA, 02138, USA
}%

\author{C. M. Knaut}
\affiliation{ 
Department of Physics , Harvard University, Cambridge, MA, 02138, USA
}%

\author{M. K. Bhaskar}
\affiliation{ 
Department of Physics , Harvard University, Cambridge, MA, 02138, USA
}%
\affiliation{ 
AWS Center for Quantum Networking, Boston, MA 02135, USA
}%

\author{A. Panday}
\affiliation{ 
John A. Paulson School of Engineering and Applied Sciences, Harvard University, Cambridge, MA, 02138, USA
}%

\author{A. M. Day}
\affiliation{ 
John A. Paulson School of Engineering and Applied Sciences, Harvard University, Cambridge, MA, 02138, USA
}%

\author{D. Renaud}
\affiliation{ 
John A. Paulson School of Engineering and Applied Sciences, Harvard University, Cambridge, MA, 02138, USA
}%

\author{M. D. Lukin}
\affiliation{ 
Department of Physics , Harvard University, Cambridge, MA, 02138, USA
}%

\author{E. Hu}
\affiliation{ 
John A. Paulson School of Engineering and Applied Sciences, Harvard University, Cambridge, MA, 02138, USA
}%

\author{B. Machielse}
\affiliation{ 
Department of Physics , Harvard University, Cambridge, MA, 02138, USA
}%
\affiliation{ 
AWS Center for Quantum Networking, Boston, MA 02135, USA
}%

\author{M. Loncar}
\affiliation{ 
John A. Paulson School of Engineering and Applied Sciences, Harvard University, Cambridge, MA, 02138, USA
}%
 \email{loncar@seas.harvard.edu}

\date{\today}

\pacs{}

\maketitle 


\begin{figure*}
\includegraphics[scale=1.25]{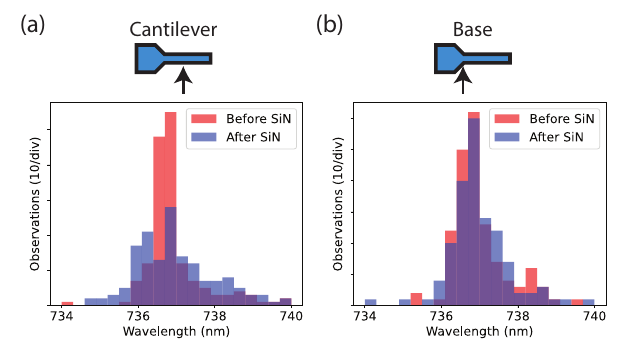}
\caption{\label{fig:simulation} Histogram of all observed SiV transition locations before and after SiN deposition, for SiVs located (a) on the cantilever or (b) on the base of the cantilever. A significant increase in the distribution of SiV optical transitions is observed upon SiN deposition in the case of SiVs on the cantilever, but not in the case of SiVs on the base. This provides further verification that strain is the origin of the broadening, as SiVs on the cantilever experience strain imparted by the thin film stressor, wheras those on the base, due to the base being directly tethered to the substrate, experience very little strain.}
\end{figure*}



%
%

%

